\documentclass[a4paper,11pt]{article}
\usepackage{pos}

\usepackage{mathtools}
\usepackage[italicdiff]{physics}
\usepackage{comment}
\usepackage{cancel}
\usepackage{booktabs}
\usepackage{feynmp-auto}
\usepackage{bm}

\newcommand{\unit}[1]{%
    \ensuremath{\,\mathrm{#1}}%
}

\newcommand{\E}[1]{%
    \times 10^{#1}%
}

\newcommand{\be}{\begin{equation}\begin{gathered}}
\newcommand{\ee}{\end{gathered}\end{equation}}

\newcommand{\bea}{\begin{equation}\begin{aligned}}
\newcommand{\eea}{\end{aligned}\end{equation}}

\title{Testing Lorentz invariance violation using cosmogenic neutrinos}
\ShortTitle{Testing LIV using cosmogenic neutrinos}

\author*[a,b]{Maykoll A. Reyes}
\author[c,d]{Denise Boncioli}
\author[a,b]{José Manuel Carmona}
\author[a,b]{José Luis Cortés}

\affiliation[a]{Departamento de Física Teórica, Universidad de Zaragoza,\\
Zaragoza, 50009, Spain}
\affiliation[b]{Centro de Astropartículas y Física de Altas Energías (CAPA),
Universidad de Zaragoza, \\
Zaragoza, 50009, Spain}
\affiliation[c]{Dipartimento di Scienze Fisiche e Chimiche, Università degli Studi dell’Aquila, \\
via Vetoio, 67100, L’Aquila, Italy}
\affiliation[d]{INFN - Laboratori Nazionali del Gran Sasso,\\
via G. Acitelli 22, 67100, Assergi (AQ), Italy}

\emailAdd{mkreyes@unizar.es}
\emailAdd{denise.boncioli@aquila.infn.it}
\emailAdd{jcarmona@unizar.es}
\emailAdd{cortes@unizar.es}

\abstract{Secondary messengers such as neutrinos and photons are expected to be produced in interactions of ultra-high-energy cosmic rays (UHECRs) with extragalactic background photons. Their propagation could be altered by the effects of Lorentz invariance violation. In this work, we have developed an extension of the SimProp code that includes some Lorentz-violating scenarios affecting the propagation of neutrinos. We present the corresponding expected cosmogenic neutrino fluxes for three different astrophysical scenarios for the production of UHECRs. These results can be used to put constraints on the scale of Lorentz violation in the neutrino sector.}
\ConferenceLogo{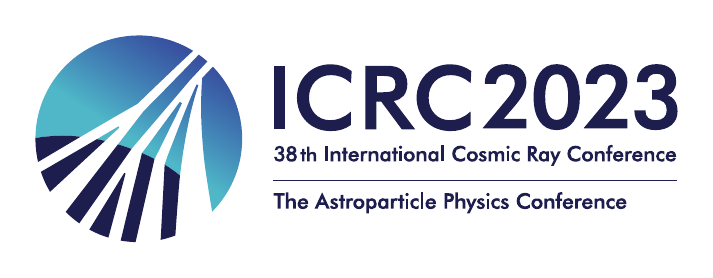}

\FullConference{%
The 38th International Cosmic Ray Conference (ICRC2023)\\
  26 July -- 3 August, 2023\\
  Nagoya, Japan}

\begin{document}
\maketitle

\section{Introduction}

Very-High-Energy Gamma Rays (VHEGRs) and Ultra-High-Energy Cosmic Rays (UHECRs) interact with the electromagnetic backgrounds in their propagation. Instead, neutrinos are very special astrophysical messengers which are only affected by the expansion of the Universe. Therefore, their observation provides a very powerful tool not only for astrophysics, but also for tests beyond the standard physics~\cite{Stecker:2022tzd}. In some quantum gravity models, the effects of Lorentz Invariance Violation (LIV) increase with the energy; consequently, cosmogenic neutrinos, produced during the propagation of UHECRs, provide one of the best playgrounds to test them.
%Cosmogenic neutrinos, produced during the propagation of UHECRs, provide one of the best playgrounds to test the effects of Lorentz Invariance Violation (LIV) increasing with the energy that appear in some models of quantum gravity.

%Neutrinos and Ultra-High-Energy Cosmic Rays (UHECRs) have been extensively studied as potential evidence for Lorentz invariance violation (LIV) (see~\cite{Stecker:2014oxa,Aloisio:2000cm}, respectively). The investigation of LIV may well require a multimessenger approach, with cosmogenic neutrinos serving as a paradigmatic example. While the maximum energies of astrophysical neutrinos and UHECRs are yet to be determined, the existence of UHECRs implies the production of cosmogenic neutrinos. In this paper, our objective is to explore the potential signals of Lorentz Invariance Violation in upcoming observations of cosmogenic neutrinos.

In this work, we introduce a LIV model affecting neutrinos within the framework of effective field theory, employing higher-dimensional operators that amplify their effects at higher energies. Specifically, we consider scenarios in which neutrinos and/or antineutrinos acquire superluminal velocities and subsequently become unstable. The decays of these particles introduce modifications to their propagation, leading to an additional energy loss mechanism alongside the energy loss due to the expansion of the universe. 
We find that the anomalies produced in the cosmogenic neutrino flux can be contrasted with the measurements of the current and close-future experiments either to find signals of LIV or to constrain the parameters of the model.
%Consequently, cosmogenic neutrinos emerge as prime candidates for detecting signals of Lorentz Invariance Violation.

\section{Theoretical framework}

We consider a LIV model (see~\cite{Carmona:2022dtp,ReyesHung:2023udr}) characterized by a rotational-invariant correction term in the neutrino free Lagrangian of order $n$ in the inverse of a scale of new physics $\Lambda$,
\be
    \mathcal{L}_{\text{free}}= 
    \bar{\nu}_{L}(i\gamma^\mu\partial_\mu)\nu_{L}-\frac{1}{\Lambda^n}\bar{\nu}_{L} \gamma^0(i\partial_0)^{n+1}\nu_{L} \,,
    \label{eq:LIV-nu}
\ee
where we have considered a negligible neutrino mass and a flavour independent correction, so that neutrino oscillations are not affected. From the free Lagrangian, one can find the Modified Dispersion Relation (MDR) for neutrinos and antineutrinos,
\be
    E =\; |\vec p| \qty[1+\alpha \qty(\frac{|\vec p|}{\Lambda})^n]\,,
\ee
with $\alpha=1$ or $(-1)^{n}$, depending on whether the particle is a neutrino or antineutrino, respectively. For the chosen sign of the LIV term in Eq.~\eqref{eq:LIV-nu}, neutrinos are always superluminal and antineutrinos will be superluminal for $n$ even. In this work we will focus on the cases $n=1$ and 2.

Due to the extra contribution in their MDR, superluminal neutrinos are unstable and can decay through the emission of an electron-positron pair, referred to as Vacuum Pair Emission (VPE), or the generation of a neutrino-antineutrino pair, known as Neutrino Splitting (NSpl).
\begin{figure}[htbp]
    \centering
    \includegraphics[width=\textwidth]{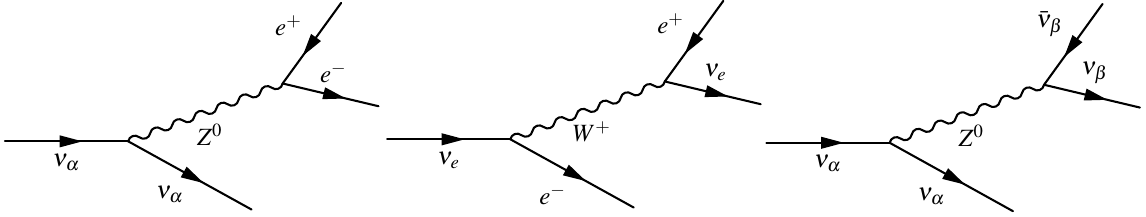}
    \caption{From left to right:\\ (a) Neutral channel of the VPE \hspace{2em}(b) Charged channel of the VPE \hspace{1em}(c) Neutral channel of the NSpl.}
    \label{fig:feynman}
\end{figure}
The VPE has a threshold given by ${E_\text{th}^{(e)} \coloneqq \qty(2 m_e^2 \Lambda^n)^{1/(2+n)}}$ and is mediated by a $Z$ boson (neutral channel, Fig.~\ref{fig:feynman}a); additionally, for electron neutrinos, it can also be mediated by a $W$ boson (charged channel, Fig.~\ref{fig:feynman}b). In contrast, the NSpl can only be mediated by a $Z$ boson (Fig.~\ref{fig:feynman}c) and has a negligible threshold due to the smallness of the neutrino mass.

Using the Feynman diagrams of Fig.~\ref{fig:feynman} and the Standard Model (SM) interaction Lagrangian, one can obtain the transition amplitude for the decay of a muon/tau or an electron neutrino through VPE and NSpl. Afterwards, one can obtain the corresponding decay widths making use of the collinearity of the decays at very high energies (as explained in~\cite{Carmona:2022dtp,ReyesHung:2023udr}), 
\begin{align}
    \Gamma_{\nu_{\mu,\tau}}^{(e)} =&\; \frac{E^5}{192\pi^3} \frac{g^4}{M_W^4} \qty[(s_W^2-1/2)^2 + (s_W^2)^2]  \qty(\frac{E}{\Lambda})^{3n} c_n^{(e)} \,, \label{eq:total_decay_muon_tau} \\
    \Gamma_{\nu_{e}}^{(e)} =&\; \frac{E^5}{192\pi^3} \frac{g^4}{M_W^4} \qty[(s_W^2-3/2)^2 + (s_W^2)^2]  \qty(\frac{E}{\Lambda})^{3n} c_n^{(e)} \,, \label{eq:total_decay_electron} \\
    \Gamma_{\nu_\alpha}^{(\nu)} =&\; 3\times \frac{E^5}{192\pi^3} \frac{g^4}{M_W^4} \qty(\frac{E}{\Lambda})^{3n} c_n^{(\nu)} \label{eq:total_decay_nu} \,,
\end{align}
where $g$ is the weak coupling, $M_W$ the mass of the $W$ boson, $s_W$ the sine of the Weinberg angle, and $c_n^{(e)}$ a constant with values $c_1^{(e)}\approx 0.14$ and $c_2^{(e)}\approx 0.18$ for $n=1$ and 2, respectively. The constant $c_n^{(\nu)}$ takes approximately the same value of $n=1$ and 2, $c_n^{(\nu)}\approx 0.024$. Similarly, one can also obtain the energy fraction probability distribution of the particles of the final state,
\begin{align}
    {\cal P}_{\nu_{\mu,\tau}}^{(e)} (x',x_-,x_+) \,=&\, \frac{1}{\,c_n^{(e)}} \,\delta(1-x'-x_+-x_-)\,(1-{x'\,}^{n+1})^3 \qty[0.61 (1-x_+)^2 + 0.39 (1-x_-)^2] \,, \label{eq:energy_dist_VPE_muon} \\
    {\cal P}_{\nu_{e}}^{(e)} (x',x_-,x_+) \,=&\, \frac{1}{\,c_n^{(e)}} \,\delta(1-x'-x_+-x_-)\,(1-{x'\,}^{n+1})^3 \qty[0.97 (1-x_+)^2 + 0.03 (1-x_-)^2] \,, \label{eq:energy_dist_VPE_electron} \\
    {\cal P}_{\nu_\alpha}^{(\nu)}(x', x_-, x_+) \,=&\, \frac{(n+1)^3}{4\,c_n^{(\nu)}} \,\delta(1-x'-x_--x_+)\,
    (1-x')^3 \,(1-x_-)^3 \, (1-x_+)^{3n-1} \label{eq:energy_dist_NSpl} \,,
\end{align}
where $x'$, $x_-$ and $x_+$ are the energy fractions of the final neutrino, and the emitted particle and antiparticle, respectively\footnote{Let us note that Eq.~\eqref{eq:energy_dist_NSpl} is symmetric in the two final neutrinos.}.

For the case of antineutrinos, if they are superluminal, they will undergo completely equivalent processes, with the same total decay width, but exchanging particles by antiparticles, and vice versa, in the Feynman diagrams and energy fraction probability distributions.

\begin{figure}[tbp]
    \centering
    \begin{minipage}{0.49\textwidth}
        \centering
        \includegraphics[width=\textwidth]{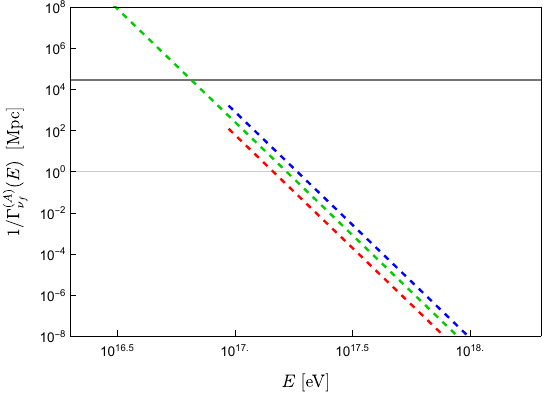}
        \caption{Decay lengths of NSpl (green), muon/tau VPE (blue) and electron VPE (red) in Mpc for $\Lambda=M_P$ and $n=2$. The VPE is only defined above $E_\text{th}^{(e)}$. The inverse of $H_0$ in shown as a black horizontal line.}
        \label{fig:decay_length}
    \end{minipage}%
    \hfill
    \begin{minipage}{0.49\textwidth}
        \centering
        \includegraphics[width=\textwidth]{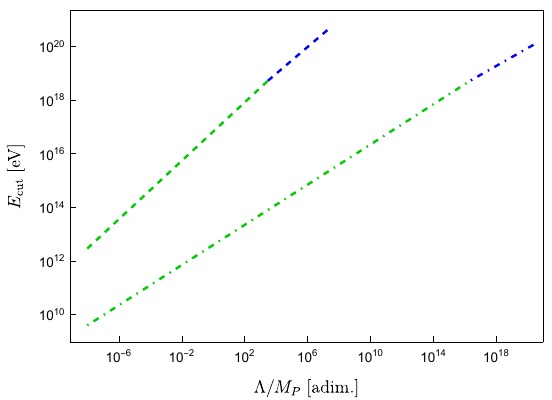}
        \caption{Approximate energy of the cutoff ($E_\text{cut}$) as a function of $\Lambda$ for $n=1$ (dash-dotted) and $n=2$ (dashed). The green part is controlled by the NSpl and the blue one by the VPE.}
        \label{fig:Ecut}
    \end{minipage}
\end{figure}

During the superluminal (anti)neutrino propagation there will be three competing processes of energy loss: the two decays (VPE and NSpl) and the expansion of the universe.
In Fig.~\ref{fig:decay_length} we show a comparison between the characteristic length of the expansion of the universe (inverse of the Hubble constant $H_0$) and the three decay lengths (inverse of decay widths), for an example scenario ($\Lambda$ equal to the Planck mass, $M_P \approx 1.2 \times 10^{28}\unit{eV}$, and $n=2$).
There, one can check how fast the decay widths grow with the energy due to the strong dependence $E^{5+3n}$. This allow us to distinguish, for each decay width, two ranges of energies: one in which $\Gamma\gg H_0$, and so the effect of the expansion is negligible with respect to the decay (instantaneous approximation), and another one with the opposite behaviour. In fact, one can define certain energy scales $E_{\nu_{\mu,\tau}}^{(e)}$, $E_{\nu_{e}}^{(e)}$, and $E_{\nu_{\alpha}}^{(\nu)}$, which act as ``effective'' or dynamical thresholds for the decays,
\be
    \Gamma_{\nu_{\mu,\tau}}^{(e)}/H_0 \doteq \qty(E/E_{\nu_{\mu,\tau}}^{(e)})^{5+3n}\,, \;\; \Gamma_{\nu_{e}}^{(e)}/H_0 \doteq \qty(E/E_{\nu_{e}}^{(e)})^{5+3n}\,, \;\;\text{and}\; \Gamma_{\nu_{\alpha}}^{(\nu)}/H_0 \doteq \qty(E/E_{\nu_{\alpha}}^{(\nu)})^{5+3n}\,.
\ee

Under the instantaneous approximation, neutrinos with energies above the effective thresholds will decay, without changing their redshift, the necessary times until falling below the thresholds. Consequentially, one expects a cutoff in the superluminal (anti)neutrino spectrum, located at the energy of the lowest (kinematical or dynamical) threshold\footnote{For (anti)neutrinos emitted from point-like sources, one should divide by $(1+z)$, where $z$ is the redshift of the source.}. In Fig.~\ref{fig:Ecut}, we show the approximate energy of the cutoff, $E_\text{cut}$, as a function of $\Lambda$.
%For values of $\Lambda<3.71\E{8} M_P$ for $n=1$ and $1.38\E{-2} M_P$ for $n=2$, the cutoff is dominated by the dynamical threshold of the NSpl; for greater values, the kinematical threshold of the VPE dominates. Taking this into account, we show the approximate location of the cutoff ($E_\text{cut}$) as a function of $\Lambda$ in Fig.~\ref{fig:Ecut}.

\section{Implementation and results of the simulations}

SimProp~\cite{Aloisio:2017iyh} is a Monte Carlo software which focuses on the propagation of cosmic rays and the produced secondary particles in their interactions with the photon backgrounds: the Cosmic Microwave Background (CMB) and the Extragalactic Background Light (EBL). We have implemented the presented LIV model in SimProp, by replacing the subroutine in charge of the neutrino propagation by a different one which takes into account the effects explained in the previous section.

%\begin{enumerate}
%    \item Check whether the particle (neutrino or antineutrino) is superluminal. If the particle is subluminal, propagate it trivially to Earth. If it is not, continue.
%    \item Check if the energy is below any threshold (kinematical or dynamical). If the energy is below any of the thresholds, set the corresponding decay width to zero. If it is not, compute the value using Eqs.~\eqref{eq:total_decay_nu}, \eqref{eq:total_decay_muon_tau} and \eqref{eq:total_decay_electron}.
%    \item If all the decay widths are zero, trivially propagate the particle to Earth. If they are not, randomly choose a process to undergo with a probability proportional to their decay widths.
%    \item Depending of the chosen effect, randomly sample the energies of the final particles from the distributions shown in Eqs.~\eqref{eq:energy_dist_NSpl}, \eqref{eq:energy_dist_VPE_muon} and \eqref{eq:energy_dist_VPE_electron} (or the corresponding ones for antineutrinos).
%\end{enumerate}

The first modification we expect with respect to Special Relativity (SR), is the production of tau neutrinos, due to the fact that the NSpl produces neutrinos of every flavour with equal probability. This will change the flavour composition at Earth, despite the fact that our LIV model does not affect neutrino oscillations. However, neutrino oscillations are not implemented in the current version of SimProp, so, in this work, we cannot make a prediction about the flavours at Earth.

Regarding the flux, the production of cosmogenic neutrinos will depend on the astrophysical scenario set for the emission of the cosmic rays from their sources. In order to identify the effects of the LIV, we consider an astrophysical scenario consisting of pure protons emitted by three possible source distributions (uniform, proportional to the Stellar Formation Rate (SFR) and proportional to the Active Galactic Nuclei (AGN) distributions~\cite{Aloisio:2015ega}) within redshift 0 and 1, and with an emission spectrum proportional to an inverse power law (with spectral indices $\gamma=2.6$, 2.5 and 2.4 for each source distribution, respectively~\cite{Aloisio:2015ega}) between $10^{17}$--$10^{21}$ eV. To incorporate the EBL into our simulations, we employed the model from~\cite{Stecker:2005qs}, available in SimProp.

In this scenario, cosmogenic neutrinos will be produced by the interactions of the protons with the photons of the CMB and EBL ($p + \gamma  \rightarrow \pi^{+}+ n$), through the decay of secondary pions ($\pi ^{+} \rightarrow \mu ^{+} + \nu_\mu$ and $\mu ^{+} \rightarrow  e^{+} + \nu _e +\bar\nu_\mu$) and from the neutron beta decay ($n \rightarrow p + e^{-} + \bar\nu_e$). If the protons interact with photons of the EBL, we expect cosmogenic neutrinos around $10^{16}$--$10^{17}\unit{eV}$; instead for photons of the CMB, the produced neutrinos will gather around $10^{18}$--$10^{19}\unit{eV}$.

\begin{figure}[tbp]
    \centering
    \begin{minipage}{0.49\textwidth}
        \centering
        \includegraphics[width=\textwidth]{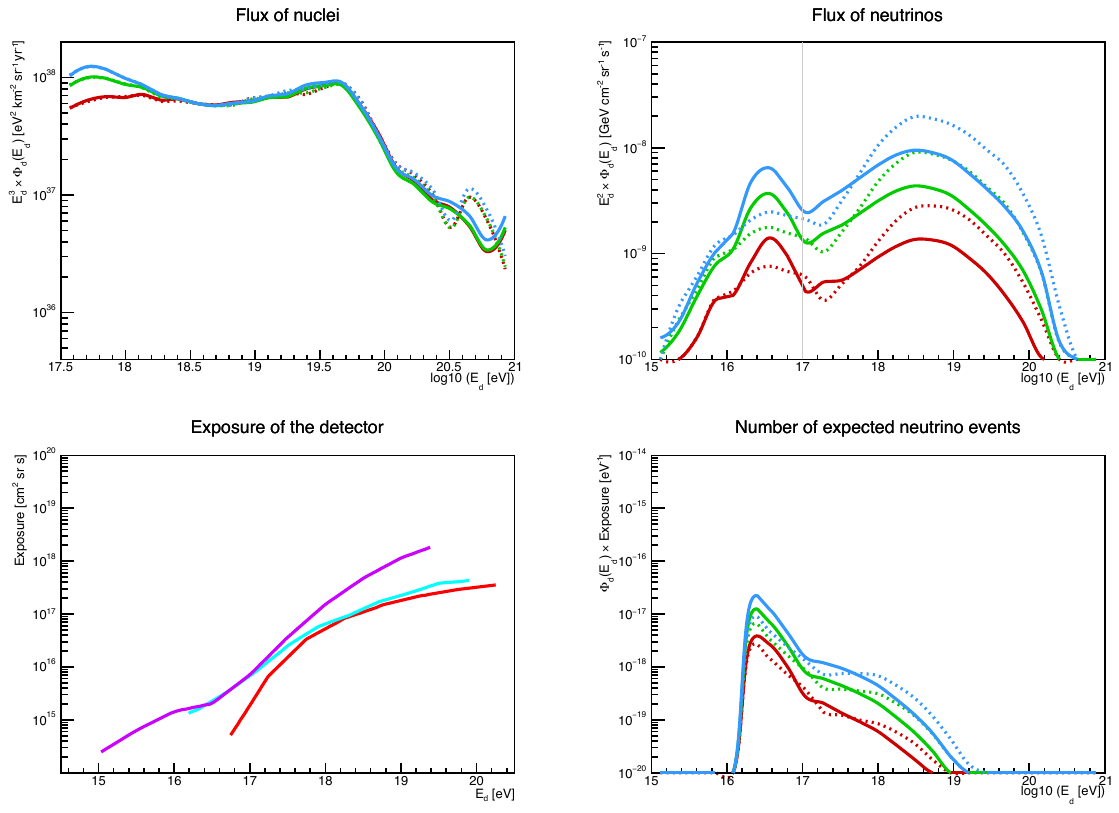}
    \end{minipage}%
    \hfill
    \begin{minipage}{0.49\textwidth}
        \centering
        \includegraphics[width=\textwidth]{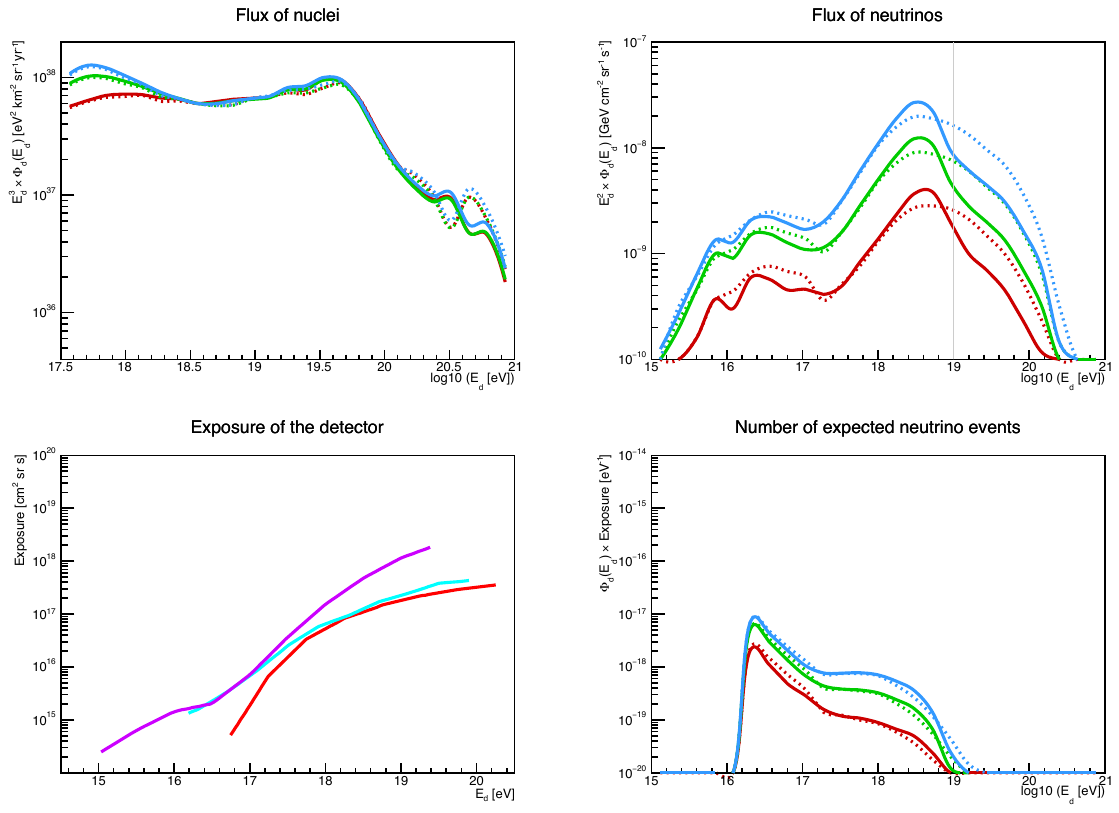}
    \end{minipage}
    \begin{minipage}{0.49\textwidth}
        \centering
        \includegraphics[width=\textwidth]{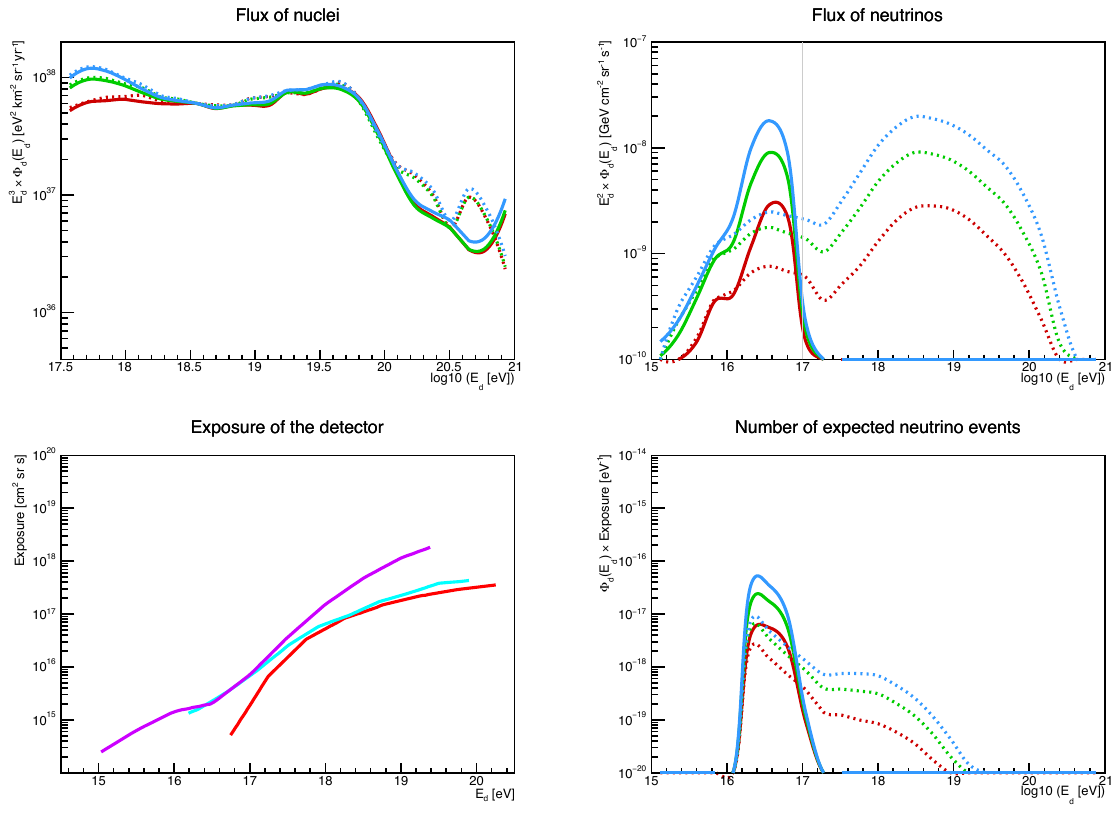}
    \end{minipage}%
    \hfill
    \begin{minipage}{0.49\textwidth}
        \centering
        \includegraphics[width=\textwidth]{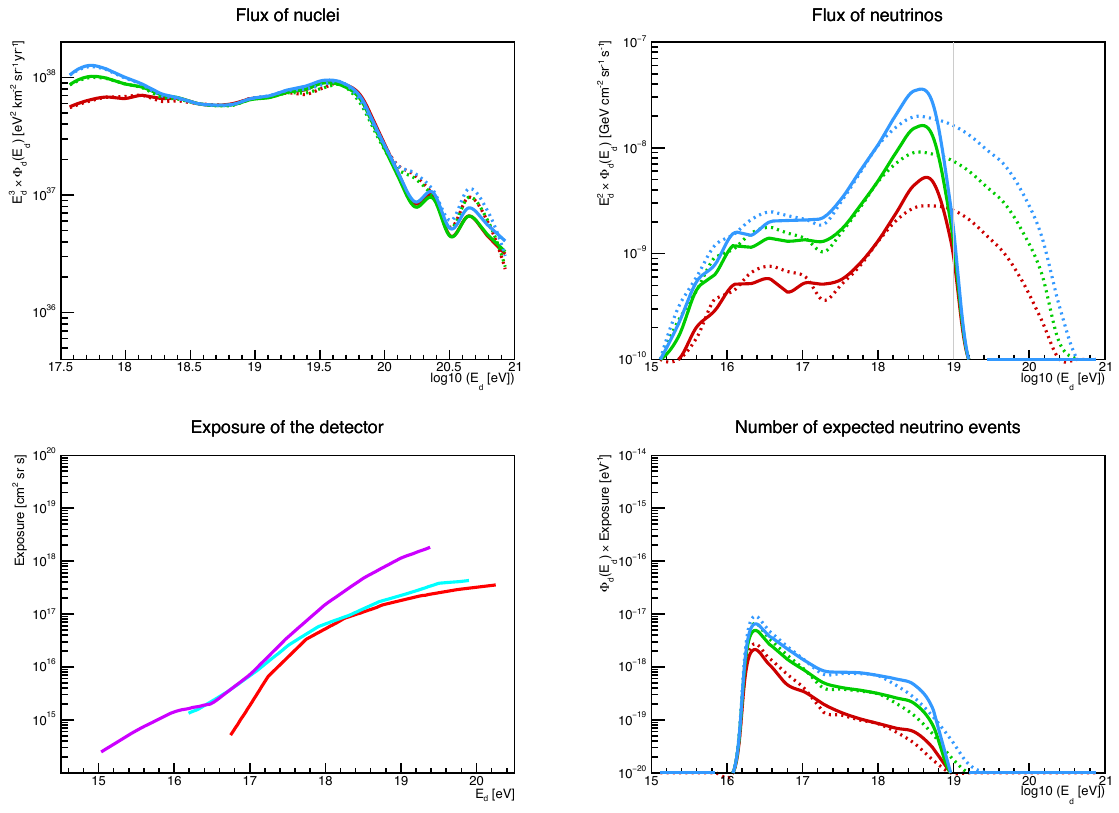}
    \end{minipage}
    \caption{Cosmogenic neutrino flux at Earth for a superluminal cutoff at  $E_\text{cut}=10^{17}\unit{eV}$ (left column) and $10^{19}\unit{eV}$ (right column), for $n=1$ (top row)  and 2 (bottom row), and for a uniform (red), SFR (green), and AGN (blue) source distribution. The corresponding SR scenario for each case are shown in dotted lines.}
    \label{fig:flux_sources}
\end{figure}

In Fig.~\ref{fig:flux_sources} we show the computed flux for the cases $n=1$ (top row, where neutrinos are superluminal and antineutrinos are subluminal) and $2$ (bottom row, where both neutrinos and antineutrinos are superluminal) and for different values of the scale of new physics $\Lambda$, such that the superluminal cutoff ($E_\text{cut}$) is around the EBL (left column) and CMB (right column) peaks. The different colors stand for the three different choices of the proton source distribution.
%\begin{figure}[htbp]
%    \centering
%    \begin{minipage}{0.49\textwidth}
%        \centering
%        \includegraphics[width=\textwidth]{img/flux_evol_n2.pdf}
%    \end{minipage}%
%    \hfill
%    \begin{minipage}{0.49\textwidth}
%        \centering
%        \includegraphics[width=\textwidth]{img/flux_evol_n1.pdf}
%    \end{minipage}
%    \caption{Cosmogenic neutrino flux at Earth for $n=2$ (left) and $n=1$ (right), for increasing values of $\Lambda$ (such that $\log_{10}(E_\text{cut}\;[\text{eV}])$ goes from 16.5 to 20 in steps of 0.5). The current Pierre Auger (red) and IceCube (cyan), and the expected 10-year IceCube Gen2 (violet), 90\% CL upper limits (from~\cite{PierreAuger:2019ens,IceCube:2018fhm,Ackermann:2022rqc}, respectively) are also shown in thick colored dashed lines.}
%    \label{fig:flux_evol}
%\end{figure}%

For the case $n=2$ (Fig.~\ref{fig:flux_sources}, bottom row), we see the expected superluminal cutoff and the existence of an anomalous excess prior to the cutoff, whose value with respect the SR curves decreases as $\Lambda$ increases. This behaviour stems from the fact that, as the value of the scale of new physics increases, so do the corresponding thresholds; in consequence, it is more difficult for neutrinos to undergo a decay and populate the \textit{bump}. One can also check that for the cases when the bump appears at energies close to the EBL peak (for $E_\text{cut}\approx 10^{17}\unit{eV}$ and $\Lambda\approx 2.19 M_P$), one gets the larger change in the flux with respect to SR; however, the maximum value of the flux (multiplied by the energy squared) is obtained when the bump appears at energies close to the CMB peak (for $E_\text{cut}\approx 10^{19}\unit{eV}$ and $\Lambda\approx 1.13\E{4} M_P$). 

For the case $n=1$ (Fig.~\ref{fig:flux_sources}, top row) one needs larger values of $\Lambda$ to get the superluminal cutoff and bump near the EBL and CMB peaks (for $E_\text{cut}\approx 10^{17}\unit{eV}$ and $10^{19}\unit{eV}$ one needs $\Lambda\approx 5.9\E{11} M_P$ and $1.57\E{17} M_P$, respectively). Additionally, due to the fact that for $n=1$ only neutrinos are superluminal, antineutrinos cannot decay, and we do not have a cutoff after $E_\text{cut}$, but just a decrease in the flux. Since the effects are not such strong with respect to the previous scenario, from now on we will focus on the case $n=2$.

\section{Current and future experimental sensitivities}

To test the sensitivities of the current and future experiments to these new physics scenarios one can compute the expected number of neutrino events, given the computed flux, using the exposure of each experiment. Currently, we have not detected neutrino events in the energy range of the cosmogenic neutrinos. Then, we can exclude at the 90\% Confidence Level (CL) all the models of LIV with a prediction in the number of expected neutrino events higher than $N=2.39$~\cite{Feldman:1997qc}.

\begin{figure}[tbp]
    \centering
    \begin{minipage}{0.49\textwidth}
        \centering
        \includegraphics[width=\textwidth]{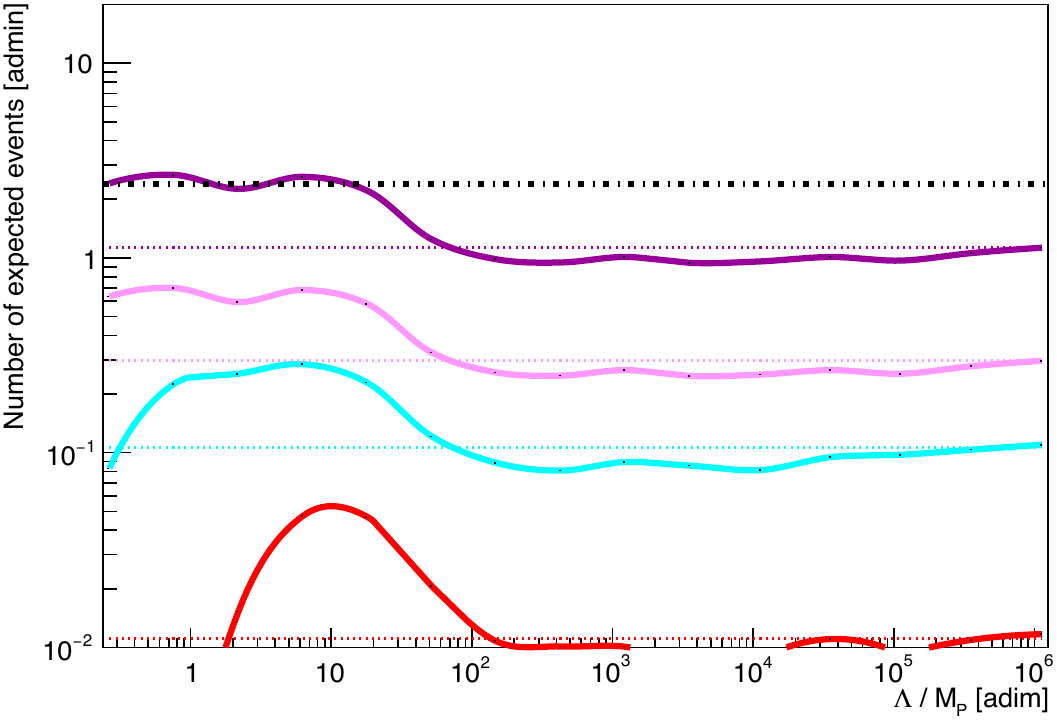}
    \end{minipage}%
    \hfill
    \begin{minipage}{0.49\textwidth}
        \centering
        \includegraphics[width=\textwidth]{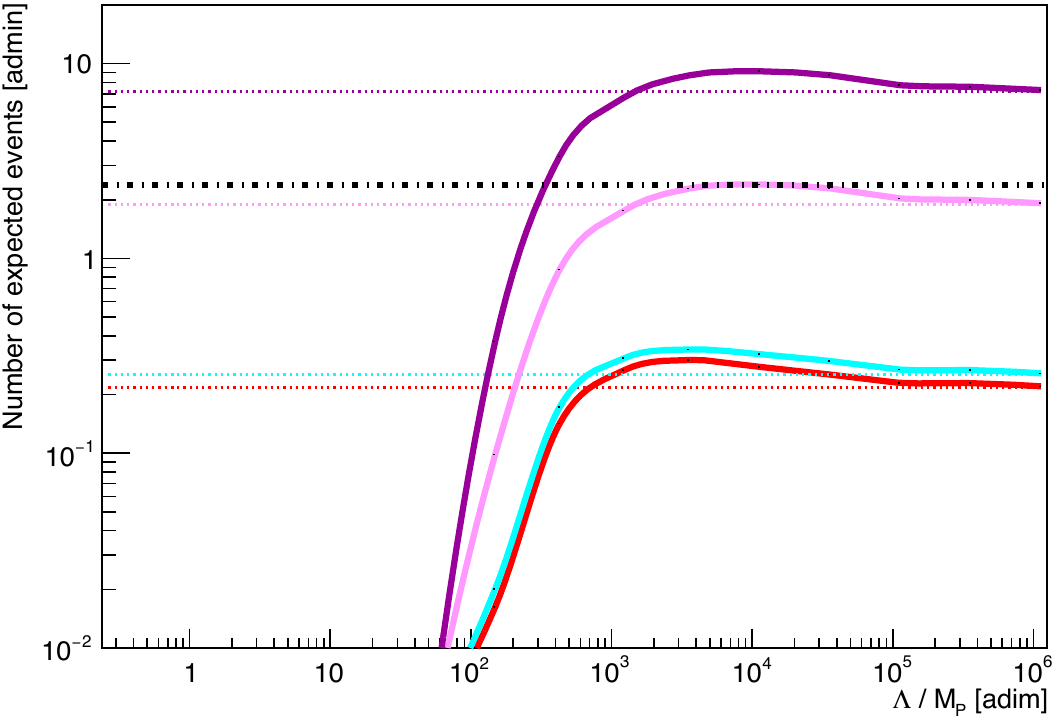}
    \end{minipage}
    \caption{Number of expected events with energies between $10^{16}$--$10^{17}$ (left) and  $10^{18}$--$10^{19}$ (right) eV by current Pierre Auger (red) and IceCube (cyan) observatories, and by a 2.1-year-old (light magenta) and 8.0-year-old (dark magenta) IceCube Gen2, with respect to the scale of new physics $\Lambda$ and for $n=2$. The corresponding SR scenario for each case is shown in dotted lines. The statistical 90\% CL UL for absence of events ($N_d=2.39$)~\cite{Feldman:1997qc} is shown in a dot-dashed black line.}
    \label{fig:events_peaks}
\end{figure}

We have extracted the current exposure of the Pierre Auger Observatory and IceCube Neutrino Observatory, and the expected exposure from IceCube Gen2 after a 2.1- and 8.0-year time window, from~\cite{PierreAuger:2019ens,IceCube:2018fhm,Ackermann:2022rqc}, respectively. Then, we computed the expected number of neutrino events, using the uniform distribution of proton sources, at the energies around the EBL ($10^{16}$--$10^{17}$ eV) and CMB ($10^{18}$--$10^{19}$ eV) peaks. The results are shown in Fig.~\ref{fig:events_peaks}, where they are compared to the statistical Upper Limit (UL) for non-observation of events.

One can see in Fig.~\ref{fig:events_peaks} that the current absence of events in Auger and IceCube cannot be used yet to put bounds in the values of the scale of new physics with the desired CL; however, if IceCube Gen2 does not detect any event between $10^{18}$--$10^{19}$ eV after 2.1 years of measurements, one would be able to reject with 90\% CL the values of $\Lambda\in[5\E{3} M_P,2\E{4} M_P]$ (Fig.~\ref{fig:events_peaks}, right). Similarly, if no neutrino event is detected between $10^{16}$--$10^{17}$ eV by IceCube Gen2 in a time window of 8.0 years, one would be able to exclude with 90\% CL the values of $\Lambda\in[2\E{-1} M_P,20 M_P]$ (Fig.~\ref{fig:events_peaks}, left). Alternatively, the non-detection of events between $10^{18}$--$10^{19}$ eV by IceCube Gen2 would favour a superluminal LIV model with $\Lambda<1.49\E{2} M_P$ and $n=2$ (i.e. with a cutoff before $E_\text{cut}\approx 10^{18}\unit{eV}$), as a possible explanation to the lack of expected events.

In a more optimistic scenario in which a non-zero flux of cosmogenic neutrinos is detected, a different statistical analysis, taking into account the uncertainties of the measured data, has to be done. However, without going into the details, some general conclusions can be drawn. For instance, the detection of neutrino events at a certain energy $E_d$ necessarily implies that, if there exists a superluminal cutoff, it must be at energies $E_\text{cut}>E_d$, which in turn can be translated into a bound on the value of $\Lambda$ (see Fig.~\ref{fig:Ecut}). At the moment, the most restrictive bound on the scale of new physics using this method comes from the recent detection of an event compatible with the Glashow resonance by IceCube~\cite{IceCube:2021rpz}, which would imply that $\Lambda>1.38\E{-2} M_P$ for $n=2$.
%and $\Lambda>3.71\E{8} M_P$ for $n=1$.
%However, future experiments like IceCube Gen2, will be able to reject some values of the parameter of new physic $\Lambda$, whose current most restrictive bound (from imposing that the cutoff must be at energies $E_\text{cut}>6.3$ PeV, due to the recent detection of an event at the Glashow resonance~\cite{IceCube:2021rpz}) are $\Lambda>1.38\E{-2} M_P$ for $n=2$ (and $\Lambda>3.71\E{8} M_P$ for $n=1$).

\section{Conclusions}

We have seen how the propagation of neutrinos can be altered by the effects of Lorentz violation. By extending the SimProp code to include such propagation effects, we have been able to present the expected cosmogenic neutrino fluxes for certain astrophysical scenarios, including a uniform, SFR, and AGN source distributions, for a pure proton composition of the UHCRs. Then, we have computed the expected number of neutrino events for current and future experiments, and compared them to the expectation for non-observation of events in the absence of expected background. We conclude that, for a uniform source distribution and a pure proton composition of the UHECR, IceCube Gen2 could be sensitive to superluminal neutrino physics in a time window of a couple of years for values of the scale $\Lambda$ giving an excess over the SR flux at energies between $10^{18}$--$10^{19}$ eV, and eight years for those values of $\Lambda$ giving an excess over the SR flux at energies between $10^{16}$--$10^{17}$ eV. Let us note that, while the flux near the CMB peak could be measured sooner than the EBL one, it would be more difficult to distinguish the standard and LIV scenarios; instead, for energies close to the EBL peak, the larger waiting time is rewarded with a larger difference between both scenarios.

The expected cosmogenic neutrino flux is also strongly influenced by the astrophysical scenario for the cosmic rays. For instance, if one includes a more realistic model with heavy nuclei as reported in~\cite{PierreAuger:2022atd}, the associated cosmogenic neutrino fluxes are smaller than those of the pure proton case. On the other hand, we are considering sources only from redshift 0 to 1; if one increases the volume of the universe under consideration, more cosmogenic neutrinos are expected, since, unlike the cosmic rays, they are not affected by the GZK cutoff. Given the limitations of this work, a more general analysis is planned for the future; however, the present study already shows the potential of cosmogenic neutrinos to put constraints on the scale of Lorentz violation in the neutrino sector.

\section*{Acknowledgments}

This work is supported by the Spanish grants PGC2022-126078NB-C21, funded by MCIN/AEI/ 10.13039/501100011033 and `ERDF A way of making Europe’, grant E21\_23R funded by the Aragon Government and the European Union, and the NextGenerationEU Recovery and Resilience Program on `Astrofísica y Física de Altas Energías’ CEFCA-CAPA-ITAINNOVA. The work of M.A.R. is supported by the FPI grant PRE2019-089024, funded by MICIU/AEI/FSE. The authors would like to acknowledge the contribution of the COST Action CA18108 ``Quantum gravity phenomenology in the multi-messenger approach''.

%\begin{thebibliography}{99}
%\bibitem{...}
%....
%\end{thebibliography}
%\bibliographystyle{apsrev}
{\setlength{\baselineskip}{1.17em}
\bibliographystyle{JHEP}
\bibliography{bib/Tesis}
}

%% Full authors list (ONLY FOR COLLABORATIONS)
%\clearpage
%\section*{Full Authors List: \Coll\ Collaboration}
%
%\noindent \textbf{Note comment afterwards:} Collaborations have the possibility to provide an authors list in xml format which will be used while generating the DOI entries making the full authors list searchable in databases like Inspire HEP. For instructions please go to icrc2021.desy.de/proceedings or contact us under icrc2021proc@desy.de.\\
%
%\scriptsize
%\noindent
%first.author$^1$, 
%second.author$^2$, 
%third.author$^3$ % .... more names
%and 
%last.author$^{n}$ \\
%
%\noindent
%$^1$first.affiliation.
%$^2$second.affiliation. % .... more affiliation
%$^{m}$last.affiliation.

\end{document}